\documentclass[aps,prl,preprint,showpacs,groupedaddress]{revtex4}
\usepackage {amssymb}
\usepackage {amsmath}
\usepackage {graphicx}
\usepackage {longtable}

\begin{document}
\title{The origin of transport barriers in toroidal plasmas}

\author{Fedor V.Prigara}
\affiliation{Institute of Microelectronics and Informatics,
Russian Academy of Sciences,\\ 21 Universitetskaya, Yaroslavl
150007, Russia}
\email{fprigara@imras.yar.ru}

\date{\today}

\begin{abstract}

A transport barrier in a toroidal plasma is treated as the
boundary region between the hot and cold phases of the plasma.
These two phases possess essentially different collision and
transport properties, so the boundary region between them
corresponds to the minimum values of specific resistivity and
transport coefficients of the plasma. The behavior of magnetic
shear in the region of a transport barrier and some other specific
properties of transport barriers are the consequences of the
minimum plasma resistivity in this region. The origin of sawtooth
oscillations in toroidal plasmas is shown to have a similar
nature.

\end{abstract}


\maketitle

Transport barriers in the tokamak plasmas, i.e. the regions of
reduced particle and energy transport across magnetic surfaces,
have been discovered as early as in 1982 [1]. There exist both
edge and internal transport barriers in the tokamak and other
toroidal plasmas. These transport barriers are normally associated
with the poloidal rotation, low (in fact, zero) magnetic shear,
and the generation of radial electric fields in the plasma [2].
With the edge transport barrier, the total energy confinement time
of the plasma in tokamaks and other toroidal devices doubles
(H-mode) over that for a plasma without the edge transport barrier
(L-mode). The theory of the classical collisional transport losses
does not explain these phenomena which are also a hard challenge
for the turbulent transport models [3].

Here we show that the transport properties of a hot plasma (with the
electron temperature exceeding the critical temperature $T_{e,c} \cong
1keV$), interacting with thermal radiation, essentially differ from those of
a cold plasma. A transport barrier in a toroidal plasma can be treated as
the boundary region between the hot and cold phases of the plasma. The
specific resistivity and the transport coefficients of the plasma have the
minimum values in this boundary region. The minimum plasma resistivity
produces a local maximum of the current density and, as a consequence, zero
magnetic shear. Some other properties of transport barriers in toroidal
plasmas are also related to the current density profile in the boundary
region.

It has been shown recently [4] that the frequency of electron-ion
collisions in a hot plasma interacting with thermal radiation is
given by the formula

\begin{equation}
\label{eq1}
\nu _{ei} = nv_{Te} \sigma _{ei} \approx nv_{Te} \sigma _{0} ,
\end{equation}

\noindent
where $n = n_{e} = n_{i} $ is the plasma density, $v_{Te} $ is the thermal
velocity of electrons, and $\sigma _{ei} $ is the cross-section of
electron-ion collisions which is a constant, $\sigma _{0} $, determined by
the atomic size. The origin of the constant cross-section $\sigma _{0} $ is
as follows. A hot plasma with the temperature $T_{e} \geqslant T_{0} \cong
3keV$ is intensely interacting with the field of thermal radiation. At
temperatures $T \geqslant T_{0} $ the stimulated radiation processes
dominate this interaction [5]. Thermal radiation induces radiative
transitions in the system of electron and ion which corresponds to the
transition of electron from the free to the bounded state.

Thus, in a hot plasma interacting with thermal radiation, the
bounded states of electrons and ions restore, leading to the
change of collision properties of a hot plasma. In this case, the
electron-ion collision cross-section has an order of magnitude of
the atomic cross-section, $\sigma _{0} \cong 10^{ - 15}cm^{2}$.

Since the specific resistivity $\eta $ of a plasma is proportional
to the frequency of electron-ion collisions, as given by the
formula

\begin{equation}
\label{eq2}
\eta = m_{e} \nu _{ei} /\left( {n_{e} e^{2}} \right),
\end{equation}

\noindent where $m_{e} $ and $e $ are the mass and charge of an
electron respectively, the resistivity of a hot plasma increases
with the electron temperature as

\begin{equation}
\label{eq3}
\eta _{H} \propto T_{e}^{1/2} ,
\end{equation}

\noindent
contrary to the relation for a cold plasma

\begin{equation}
\label{eq4}
\eta _{C} \propto T_{e}^{ - 3/2} .
\end{equation}

The critical value $T_{e,c} $ of the electron temperature
corresponding to the transition from the cold to the hot phase of
a plasma has an order of magnitude of the inversion temperature
$T_{0} \cong 3keV$ [5]. The studies of sawtooth oscillations by
von Goeler, Stodiek, and Sauthoff [6] suggest that the critical
temperature is approximately $T_{e,c} \cong 0.8keV$ (see below).

For a typical high-density discharge in a tokamak, the electron temperature
is monotonously decreasing with the radius from the center of the plasma to
the periphery. If the electron temperature at the center of the plasma,
$T_{e,0} $, exceeds the critical value, $T_{e,c} $, then the region of the
minimum resistivity exists, as it follows from the relations (\ref{eq3}) and (\ref{eq4}).
The current density tends to concentrate in this region of low resistivity.

The diffusion coefficients, $D_{ \bot}  $, and thermal conductivity, $\chi
_{ \bot}  $, describing the transport of particles and energy across the
magnetic surfaces, are proportional to the Larmor radius, $\rho $, squared
and to the frequency of electron-ion collisions:

\begin{equation}
\label{eq5}
\chi _{e, \bot}  \propto \rho _{e}^{2} \nu _{ei} ;
\quad
\chi _{i, \bot}  \propto \rho _{i}^{2} \nu _{ei} ,
\end{equation}

\noindent
for electrons and ions respectively.

The electron Larmor radius is

\begin{equation}
\label{eq6}
\rho _{e} = c\left( {m_{e} T_{e}}  \right)^{1/2}/\left( {eB} \right),
\end{equation}

\noindent
where \textit{B} is the magnetic field strength, and \textit{c} is the speed
of light, so $\chi _{e, \bot}  \propto T_{e}^{3/2} $ in the hot phase, and
$\chi _{e, \bot}  \propto T_{e}^{ - 1/2} $ in the cold phase. Similar is the
temperature dependence of the ion thermal conductivity.

Therefore, in the boundary region between the hot and cold phases
of the plasma, both the specific resistivity and transport
coefficients of the plasma have the minimum values. These regions
of reduced particle and energy transport across magnetic surfaces
have been indeed observed experimentally and called transport
barriers. There exist edge and internal transport barriers in
toroidal plasmas. The edge transport barrier corresponds to the
hot phase of the plasma occupying almost all plasma column which
has been symbolically called H-mode (here \textit{H} originally
denoted the confinement of the plasma, not the hot phase).

Returning to the resistivity profile of a toroidal plasma, we note
that the local maximum of the current density, \textit{J}, is
associated with a transport barrier. It implies that in this
region the derivative of the current density with respect to the
radius, \textit{r}, is zero, i.e. $dJ/dr = 0$. The safety factor,
\textit{q}, is defined by the formula

\begin{equation}
\label{eq7}
q\left( {r} \right) = B_{T} r/B_{P} R,
\end{equation}

\noindent
where $B_{T} $ is the toroidal magnetic field, \textit{R} is the major
radius of the plasma, and

\begin{equation}
\label{eq8}
B_{P} = 2\pi \langle J\rangle r/c
\end{equation}

\noindent
is the poloidal magnetic field produced by the toroidal current in the
plasma, $\langle J\rangle $ being the current density averaged over the
plasma column within the radius \textit{r}. Substituting expression (\ref{eq8}) in
the equation (\ref{eq7}), we obtain

\begin{equation}
\label{eq9}
q\left( {r} \right) = cB_{T} /\left( {2\pi \langle J\rangle R} \right).
\end{equation}

Now if the current density is locally constant, $dJ/dr = 0$, then the
averaged current density is also locally constant, and we find that the
magnetic shear, \textit{S}, defined by the formula

\begin{equation}
\label{eq10}
S = \left( {r/q} \right)dq/dr
\end{equation}

\noindent
is zero in the region of a transport barrier.

Exactly such a relation between transport barriers and the regions
of low magnetic shear is observed experimentally in toroidal
plasmas [2]. The additional current density parallel to the
magnetic field in the region of a transport barrier and caused by
the local minimum of plasma resistivity is responsible both for
the magneto-hydrodynamic activity associated with the moving
region of zero magnetic shear, and for the relation between the
transport barriers and the rational magnetic surfaces with $q =
m/n$ for low \textit{m} and \textit{n} [7].

When the electron temperature at the center of the plasma column
only slightly exceeds the critical value $T_{e,c} $, the sawtooth
oscillations occur [6]. The sawtooth oscillations or internal
disruptions have been observed both in high-density discharges
near the high-density instability limit and at low densities near
the electron runaway regime. It suggests that it is the value of
the electron temperature which is crucial for the development of
sawtooth oscillations, but not the electron density. The sawtooth
oscillations occur only in the central region of the plasma column
where the electron temperature is sufficiently high.

During the heating of the plasma, the plasma temperature profile
in the plasma core sharpens until the region of the minimum plasma
resistivity is formed, in accordance with above considerations. As
a consequence, the current density profile and then the electron
temperature profile flatten, causing a decrease inside and
increase just outside the $q = 1$ surface [6], indicating that the
region of low plasma resistivity is connected to the rational $q =
1$magnetic surface. The latter connection explains the development
of the sinusoidal $m = 1,n = 1$ kink mode associated with the
sawtooth oscillation [6].

To summerise, we show that the formation of a transport barrier in
a toroidal plasma is caused by the minimum plasma resistivity and
transport coefficients in the boundary region between the hot and
cold phases of the plasma. The additional current density in the
boundary region caused by the minimum plasma resistivity is
responsible for the connection between the transport barriers and
the rational magnetic surfaces. There exists a critical value of
the electron temperature corresponding to the transition from the
cold to the hot phase of the plasma. At some conditions, the
sawtooth oscillations develop in the boundary region between the
cold and hot phases of the plasma.

\begin{center}
---------------------------------------------------------------
\end{center}

[1] F.Wagner et al. Phys. Rev. Lett. \textbf{49}, 1408 (1982).

[2] K.A.Razumova, Usp. Fiz. Nauk \textbf{171}, 329 (2001)
[Physics-Uspekhi \textbf{44}, 311 (2001)].

[3] A.C.C.Sips, Plasma Phys. Control. Fusion \textbf{47}, A19
(2005).

[4] F.V.Prigara, Ann. Phys. (submitted), E-print archives, physics/0410102.

[5] F.V.Prigara, in Gamow Memorial International Conference, 8-14
August 2004, Odessa, Ukraine, E-print archives, astro-ph/0310491.

[6] S.von Goeler, W.Stodiek, and N.Sauthoff, Phys. Rev. Lett. \textbf{33},
1201 (1974).

[7] K.A.Razumova et al. Plasma Phys. Control. Fusion \textbf{42}, 973
(2000).

\end{document}